\documentclass[aps,twocolumn,showpacs,byrevtex,prl,reprint]{revtex4-1}
\usepackage{epsfig}
\usepackage{graphicx}% Include figure files
\usepackage{dcolumn}% Align table columns on decimal point
\usepackage{bm}% bold math
\usepackage{overpic}
\usepackage{subfigure}
\usepackage{float}
\usepackage{color}
\usepackage{amsmath}
\usepackage{mathcomp}
\usepackage{mathrsfs}
\usepackage{multirow}
\usepackage{rotating}

\begin{document}
\normalsize
\parskip=5pt plus 1pt minus 1pt

%\linenumbers
\title{ \boldmath Amplitude analysis of $D_{s}^{+} \rightarrow \pi^{+}\pi^{0}\eta$  
and first observation of the pure $W$-annihilation decays $D_{s}^{+} \rightarrow a_{0}(980)^{+}\pi^{0}$ and $D_{s}^{+} \rightarrow a_{0}(980)^{0}\pi^{+}$}
\vspace{-1cm}

\author{
   \begin{small}
    \begin{center}
      M.~Ablikim$^{1}$, M.~N.~Achasov$^{10,d}$, S. ~Ahmed$^{15}$, M.~Albrecht$^{4}$, M.~Alekseev$^{56A,56C}$, A.~Amoroso$^{56A,56C}$, F.~F.~An$^{1}$, Q.~An$^{53,43}$, J.~Z.~Bai$^{1}$, Y.~Bai$^{42}$
% ...
         \\
         \vspace{0.2cm}
   (BESIII Collaboration)\\
\vspace{0.2cm} {\it
$^{1}$ Institute of High Energy Physics, Beijing 100049, People's Republic of China\\
% ...
$^{4}$ Bochum Ruhr-University, D-44780 Bochum, Germany\\
% ...
$^{10}$ G.I. Budker Institute of Nuclear Physics SB RAS (BINP), Novosibirsk 630090, Russia\\
% ...
$^{15}$ Helmholtz Institute Mainz, Johann-Joachim-Becher-Weg 45, D-55099 Mainz, Germany\\
% ...
$^{42}$ Southeast University, Nanjing 211100, People's Republic of China\\
$^{43}$ State Key Laboratory of Particle Detection and Electronics, Beijing 100049, Hefei 230026, People's Republic of China\\
% ...
$^{53}$ University of Science and Technology of China, Hefei 230026, People's Republic of China\\
% ...
$^{56}$ (A)University of Turin, I-10125, Turin, Italy; (B)University of Eastern Piedmont, I-15121, Alessandria, Italy; (C)INFN, I-10125, Turin, Italy\\
% ...
$^{d}$ Also at the Novosibirsk State University, Novosibirsk, 630090, Russia\\
% ...
}\end{center}
\vspace{0.4cm}
\end{small}
}

\affiliation{}
\vspace{-4cm}
\date{\today}

\begin{abstract}
We present the first amplitude analysis of the decay $D^{+}_{s} \rightarrow \pi^{+}\pi^{0}\eta$. We 
use an $e^{+}e^{-}$ collision data sample corresponding to an integrated luminosity of 3.19~${\mbox{\,fb}^{-1}}$ 
collected with the BESIII detector at a center-of-mass energy of $4.178$ GeV. 
%The pure $W$-annihilation decay $D_{s}^{+} \rightarrow a_{0}(980)\pi$ is observed for the first time with a 
%statistical significance greater than 16 standard deviations. 
We observe for the first time the pure $W$-annihilation decays $D_{s}^{+} \rightarrow a_{0}(980)^{+}\pi^{0}$ and $D_{s}^{+} \rightarrow a_{0}(980)^{0}\pi^{+}$.  
We measure the absolute branching fractions 
$\mathcal{B}(D_{s}^{+} \rightarrow a_{0}(980)^{+(0)}\pi^{0^(+)}, a_{0}(980)^{+(0)} \to \pi^{+(0)}\eta) = (1.46\pm0.15_{{\rm stat.}}\pm0.23_{{\rm sys.}})$\%,
which is larger than the branching fractions of other measured pure $W$-annihilation decays 
by at least one order of magnitude.  
In addition, we measure the branching fraction of $D_{s}^{+} \rightarrow \pi^{+}\pi^{0}\eta$ 
with significantly improved precision. 

\end{abstract}
\pacs{13.25.Ft, 12.38.Qk, 14.40.Lb}
\maketitle
The theoretical understanding of the weak decay of charm mesons is challenging because the charm quark mass 
is not heavy enough to describe exclusive processes with a heavy-quark expansion. 
The $W$-annihilation (WA) process may occur as a result of  
final-state-interactions (FSI) and the WA amplitude 
may be comparable with the tree-external-emission amplitude~\cite{HaiYangCheng1,Li:2012cfa,Li:2013xsa,HaiYangCheng2}.   
However, the theoretical calculation of the WA amplitude is currently difficult. 
Hence measurements of decays involving a WA contribution 
provide the best method to investigate this mechanism. 

Among the measured decays involving WA contributions, two decays with $VP$ mode,    
$D_{s}^{+} \rightarrow \omega \pi^{+}$ and $D_{s}^{+} \rightarrow \rho^{0} \pi^{+}$,  
only occur through WA, which we refer to as `pure WA decay'. Here 
$V$ and $P$ denote vector and pseudoscalar mesons, respectively. 
The branching fractions (BFs) of these pure WA decays are at the $\mathcal{O}(0.1\%)$~\cite{PDG}. 
These BF measurements allow the determination of two distinct WA amplitudes for $VP$ mode. 
In addition, they improve our understanding of SU(3)-flavor symmetry and $CP$ violation in the charm sector~\cite{HaiYangCheng2}. 
However, for $SP$ mode, where $S$ denotes a scalar meson, there are neither experimental measurements nor theoretical calculations of the BFs.

\begin{figure}[htbp]
\begin{center}
\begin{minipage}[b]{0.235\textwidth}
\epsfig{width=0.98\textwidth,clip=true,file=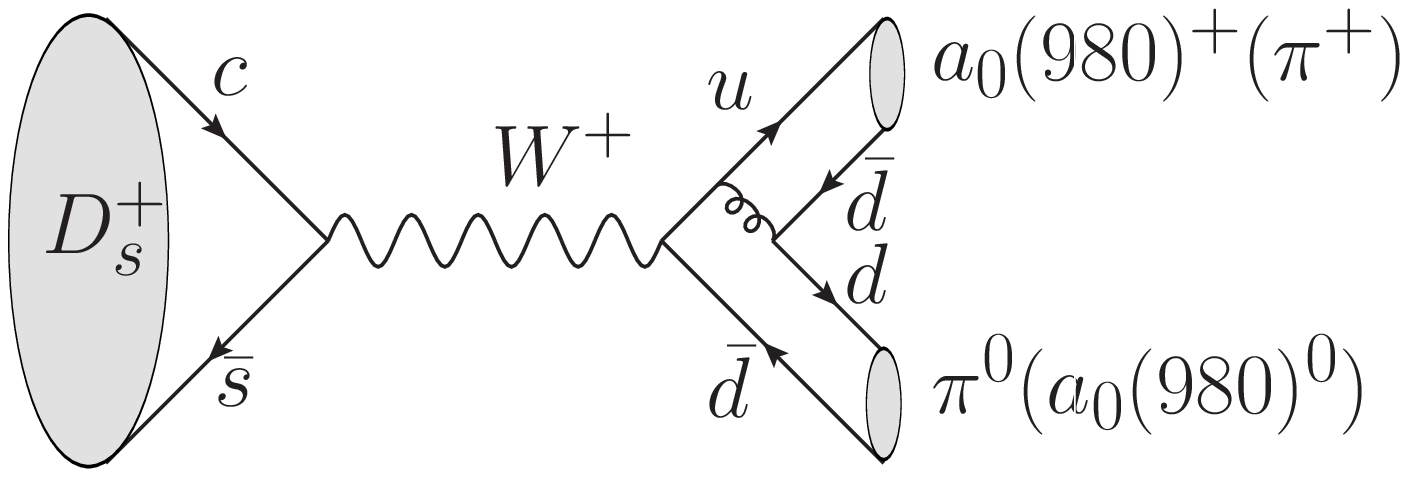}
\end{minipage}
\begin{minipage}[b]{0.235\textwidth}
\epsfig{width=0.98\textwidth,clip=true,file=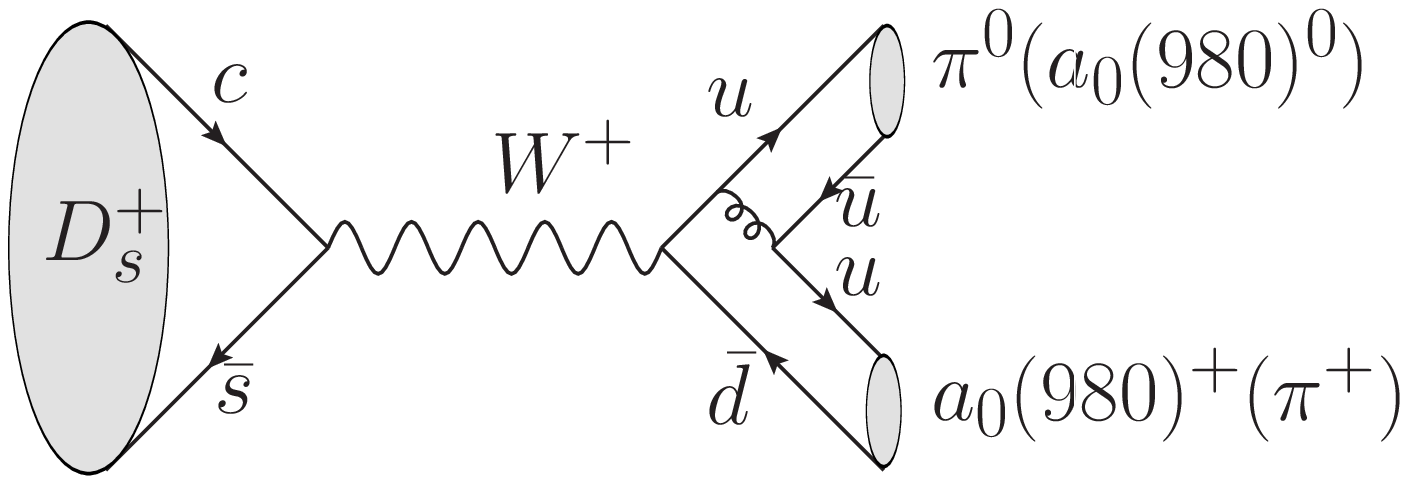}
\end{minipage}
\caption{$D_{s}^{+} \rightarrow a_{0}(980)^{+(0)}\pi^{0(+)}$ decay topology diagrams, where 
the gluon lines can be connected with the quark lines in all possible cases and 
the contributions from FSI are included. 
}
\label{fig:A}
\end{center}
\end{figure}

Two decays with $SP$ mode $D_{s}^{+} \rightarrow a_{0}(980)^{+}\pi^{0}$
and $D_{s}^{+} \rightarrow a_{0}(980)^{0}\pi^{+}$ are pure WA decays if $a_{0}(980)^{0}$-$f_{0}(980)$ mixing is ignored.
Their decay diagrams are shown in Fig.~\ref{fig:A}. 
In this Letter, we search for them with an amplitude analysis of $D_{s}^{+} \rightarrow \pi^{+}\pi^{0}\eta$. 
We also present improved measurements of the BFs of
$D_{s}^{+} \rightarrow \pi^{+}\pi^{0}\eta$ and $D_{s}^{+} \rightarrow \rho^{+}\eta$ decays. 
Throughout this Letter, charge conjugation and $a_{0}(980) \rightarrow \pi \eta$ are implied unless explicitly stated. 

We use a data sample
corresponding to an integrated luminosity of 3.19 ${\mbox{\,fb}^{-1}}$, taken at a center-of-mass 
energy of $4.178$~GeV with the BESIII detector located at Beijing Electron 
Position Collider~\cite{Yu:IPAC2016-TUYA01}. 
The BESIII detector and the upgraded multi-gap resistive plate chambers used in the time-of-flight systems are described in Refs.~\cite{detector} and \cite{MRPC}, respectively.
We study the background and determine tagging efficiencies with a generic Monte Carlo (GMC) sample that is simulated with {\sc geant4}~\cite{sim}. The GMC sample includes all known open-charm decay processes, which are generated with {\sc conexc}~\cite{ConExc} and {\sc evtgen}~\cite{EvtGen}, initial-state radiative decays to the $J/\psi$ or $\psi(3686)$, and continuum processes. 
We determine signal efficiencies from Monte Carlo (MC) samples of $D_{s}^{+} \rightarrow \pi^{+}\pi^{0}\eta$ decays that are generated according to the amplitude fit results to data reported in this Letter.

In the data sample, the $D_{s}$ mesons are mainly produced via the process of $e^{+}e^{-} \rightarrow D_{s}^{*-}D_{s}^{+}$, 
$D_{s}^{*-}\rightarrow \gamma D_{s}^{-}$; we refer to the $\gamma$ directly produced from the $D^{*-}_{s}$ decay as $\gamma_{\rm direct}$. 
To exploit the dominance of the $e^{+}e^{-} \rightarrow D_{s}^{*-}D_{s}^{+}$ process, we use the double-tag (DT) method~\cite{tagmethod}. 
The single-tag (ST) $D_s^{-}$ mesons are reconstructed using seven hadronic decays: 
$D_{s}^{-} \rightarrow K_{S}^{0} K^{-}$, $D_{s}^{-} \rightarrow K^{+} K^{-} \pi^{-}$,
$D_{s}^{-} \rightarrow K_{S}^{0} K^{-} \pi^{0}$, $D_{s}^{-} \rightarrow K^{+} K^{-} \pi^{-} \pi^{0}$,
$D_{s}^{-} \rightarrow K_{S}^{0} K^{+} \pi^{-}\pi^{-}$, $D_{s}^{-} \rightarrow \pi^{-}\eta$, and
$D_{s}^{-} \rightarrow \pi^{-}\eta^{\prime}$. 
A DT is formed by selecting a $D_{s}^{+} \to \pi \pi^{0} \eta$ decay in the side of the event recoiling against the $D_{s}^{-}$ tag. 
Here, $K_{S}^{0}$, $\pi^{0}$, $\eta$ and $\eta^{\prime}$ 
are reconstructed using $\pi^{+}\pi^{-}$, $\gamma\gamma$, $\gamma\gamma$ and $\pi^{+}\pi^{-}\eta$ channels, respectively. 
The selection criteria for charged tracks, photons, $K_{S}^{0}$ and $\pi^{0}$ are the same as those reported in Ref.~\cite{omegapi}. 
The $\eta^{(\prime)}$ candidate is required to have an invariant mass of 
the $\gamma\gamma$ $(\pi^{+}\pi^{-}\eta)$ 
combination in the interval $[0.490,\,0.580]~([0.938,\,0.978])$~GeV$/c^{2}$.  

The invariant mass of the tagged (signal) $D_s^{-(+)}$ candidates $M_{\rm tag}~(M_{\rm sig})$ are required to be in the interval $[1.90,\,2.03]$ GeV/$c^2$~($[1.87,\,2.06]$ GeV/$c^2$). 
For the ST $D_{s}^{-}$ mesons, the recoil mass 
$M_{{\rm rec}} = \sqrt{(E_{{\rm tot}}-\sqrt{|\vec{\bf p}_{D_{s}}|^{2}+m_{D_{s}}^{2}})^{2} 
        - |\vec{\bf p}_{{\rm tot}} - \vec{\bf p}_{D_{s}}|^{2}}$ is required to be within the range 
$[2.05,\,2.18]$ GeV$/c^{2}$ to suppress events from non-$D_{s}^{*-}D_{s}^{+}$ processes. Here, $(E_{{\rm tot}},\vec {\bf p}_{{\rm tot}})$ is the four-momentum of the colliding $e^{+}e^{-}$ system,  
$\vec{\bf p}_{D_{s}}$ is the three-momentum of the $D_{s}$ candidate and $m_{D_{s}}$ is the $D_{s}$ mass~\cite{PDG}. 
For events with multiple tag candidates for a single tag mode, the one with a value of $M_{{\rm rec}}$ 
closest to $m_{D_{s}}$ is chosen. 
If there are multiple signal candidates present against a selected tag candidate, 
the one with a value of $(M_{\rm tag}+M_{\rm sig})/2$ closest to $m_{D_{s}}$ is accepted. 

We perform a seven-constraint (7C) kinematic fit to the selected DT candidates for two reasons. 
First, to successfully perform an amplitude analysis, the 7C fit ensures that all events fall within the Dalitz plot. Second, it allows the selection of the $\gamma_{\rm direct}$ candidate.
In the 7C fit, aside from constraints arising from four-momentum conservation, the invariant masses of the $(\gamma\gamma)_{\pi^{0}}$, $(\gamma\gamma)_{\eta}$, 
and $\pi^{+}\pi^{0}\eta$ combinations used to reconstruct the signal $D^{+}_{s}$ candidate are constrained to the nominal $\pi^{0}$, $\eta$  and $D^{+}_{s}$ masses \cite{PDG}, respectively. The $\gamma_{\rm direct}$ candidate used in the 7C fit that produces the smallest $\chi^{2}_{\rm {7C}}$ is selected. We require $\chi^{2}_{\rm {7C}}<1000$ to avoid introducing a broad peak in the $M_{{\rm sig}}$ background distribution arising from events that are inconsistent with the signal hypothesis. 
To further suppress the background, we perform another 7C kinematic 
fit, referred to as the `7CA fit', by replacing the signal $D_{s}^{+}$ mass constraint with a $D_{s}^{*}$ mass constraint in which the invariant mass of 
either the $D_{s}^{+}$ or $D_{s}^{-}$ candidate and the selected $\gamma_{{\rm direct}}$ is constrained to the nominal $D_{s}^{*}$  mass \cite{PDG}. 
We require one of the values of the $\chi^{2}_{{\rm 7CA}}$ to be less than 500, to ensure reasonable consistency with the signal hypothesis. 
To suppress the background associated with the fake $\gamma_{\rm direct}$ candidates in the signal events, 
we veto events with $\cos \theta_{\eta} < 0.998$, where $\theta_{\eta}$ 
is the angle between the $\eta$ momentum vector from a $\eta$ mass constraint fit 
and that from the 7CA kinematic fit.
After applying these criteria, we further reduce the background, by using a multi-variable analysis method~\cite{Hocker:2007ht} in which a boosted decision tree (BDT) classifier is developed using the GMC sample. The BDT takes three discriminating variables as inputs: the invariant mass of the photon pair used to reconstruct the $\eta$ candidate, the momentum of the lower-energy photon from the $\eta$ candidate, and the momentum of the $\gamma_{{\rm direct}}$ candidate.
Studies of the GMC sample show that a requirement on the output of the 
BDT retains 77.8\% signal and rejects 73.4\% background. 
Events in which the signal candidate lies within the interval $1.93<M_{{\rm sig}}<1.99$~GeV$/c^{2}$ are retained for the amplitude analysis.  
The background events in the signal region from the GMC sample are used to model the corresponding background in data. To check the accuracy of the GMC background modeling,  we compare the $M_{\pi^{-}\pi^{0}}$, $M_{\pi^{+}\eta}$ and $M_{\pi^{0}\eta}$ distributions of events outside the selected $M_{\rm sig}$ interval between data and the GMC sample; the distributions are found to be compatible within the statistical uncertainties. 
We retain a sample of 1239 $D^{+}_{s} \to \pi^{+}\pi^{-}\eta$ candidates that has a purity of $(97.7\pm0.5)$\%.

The amplitude analysis is performed using an unbinned maximum-likelihood fit to the accepted candidate events in data. 
The background contribution is subtracted in the likelihood calculation
by assigning negative weights to the background events. 
The total amplitude $\mathcal{M}(p_{j})$ is modeled as the coherent sum of the amplitudes of all intermediate processes,
$\mathcal{M}(p_{j}) = \sum{c_{n}e^{i\phi_{n}}A_{n}(p_{j})}$, where $c_{n}$ and $\phi_{n}$ are the 
magnitude and phase of the $n^{{\rm th}}$ amplitude, respectively.
The $n^{{\rm th}}$ amplitude $A_{n}(p_{j})$ is given by 
$A_{n}(p_{j}) = P_{n}S_{n}F_{n}^{r}F_{n}^{D}$.  
Here $P_{n}$ is a function that describes the propagator of the intermediate resonance. The resonance $\rho^{+}$
is parameterized by a relativistic Breit-Wigner function, while the resonance $a_{0}(980)$ is parameterized as a 
two-channel-coupled Flatt\'e formula ($\pi \eta$ and $K\bar{K}$), 
$P_{a_{0}(980)} = 
1/((m^{2}_{0} - s_{a})-i(g^{2}_{\eta\pi}\rho_{\eta\pi}+g^{2}_{K\bar{K}}\rho_{K\bar{K}}))$.   
Here, $\rho_{\eta\pi}$ and $\rho_{K\bar{K}}$ are the phase space factors: $2q/\sqrt{s_{a}}$, 
where $q$ is denoted as the magnitude of the momentum of the daughter particle 
in the rest system and $s_{a}$ is the invariant mass square of $a_{0}(980)$.
We use the coupling constants $g^{2}_{\eta\pi} = 0.341\pm0.004$ GeV$^{2}/c^{4}$ 
and $g^{2}_{K\bar{K}} = (0.892\pm0.022)g^{2}_{\eta\pi}$, reported in Ref.~\cite{BAM-168}.
The function $S_{n}$ describes angular-momentum conservation in the decay and is constructed using the covariant tensor formalism~\cite{Zou}. 
The function $F_{n}^{r(D)}$ is the Blatt-Weisskopf barrier factor of the intermediate state ($D_{s}$ meson).
Further, according to the topology diagrams shown in Fig.~\ref{fig:A}, the W-annihilation amplitudes of decays $D_{s}^{+} \rightarrow a_{0}(980)^{+}\pi^{0}$ 
and $D_{s}^{+} \rightarrow a_{0}(980)^{0}\pi^{+}$ imply the relationship 
$A(D_{s}^{+} \rightarrow a_{0}(980)^{+}\pi^{0}) = - A(D_{s}^{+} \rightarrow a_{0}(980)^{0}\pi^{+})$. 

For each amplitude, the statistical significance is determined from the change in $2\ln\!\mathcal{L}$ and the number of degrees of freedom (NDOF) when the fit is performed with and without the amplitude included. 
In the nominal fit, only amplitudes that have a significance greater than $5\sigma$ are considered, where $\sigma$ is the standard deviation. 
In addition to the $D^{+}_{s}\to \rho^{+}\eta$ amplitude, both $D_{s}^{+} \rightarrow a_{0}(980)^{+}\pi^{0}$ and
$D_{s}^{+} \rightarrow a_{0}(980)^{0}\pi^{+}$ amplitudes are found to be significant. 
However, the latter two amplitude phases are found to be approximately $90\%$ correlated with one another;
their fitted $c_{n}$ are found to be consistent with each other while a difference in $\phi_n$ is found to be close to $\pi$, 
which indicates there is no significant $a_{0}(980)^{0} - f_{0}(980)$ mixing in $D_{s}^{+} \rightarrow a_{0}(980)^{0}\pi^{+}$.  
Consequently, in the nominal fit, we set the values of $c_n$ of these two amplitudes to be equal  with a phase difference of $\pi$. We refer to the coherent sum of these two amplitudes as ``$D_{s}^{+} \rightarrow a_{0}(980)\pi$''. 
The non-resonant process $D_{s}^{+} \rightarrow (\pi^{+}\pi^{0})_{V}\eta$ 
is also considered, where the subscript $V$ denotes a vector non-resonant state of the $\pi^{+}\pi^{0}$ combination. 
We consider other possible amplitudes that involve $\rho(1450)$, $a_{0}(1450)$, $\pi_{1}(1400)$, $a_{2}(1320)$, or $a_{2}(1700)$, 
as well as the 
non-resonant partners; none of these amplitudes has a statistical significance greater than $2\sigma$, so they are not included in the nominal model.  
In the fit, the values of $c_n$ and $\phi_n$ for the $D_{s}^{+} \rightarrow \rho^{+} \eta$ 
amplitude are fixed to be one and zero, respectively, so that all other amplitudes are measured relative to this amplitude. The masses and widths of the intermediate resonances used in the fit, except for those of the $a_{0}(980)$, are 
taken from Ref.~\cite{PDG}.

\begin{figure}[htp]
\begin{center}
\begin{minipage}[b]{0.23\textwidth}
\epsfig{width=0.98\textwidth,clip=true,file=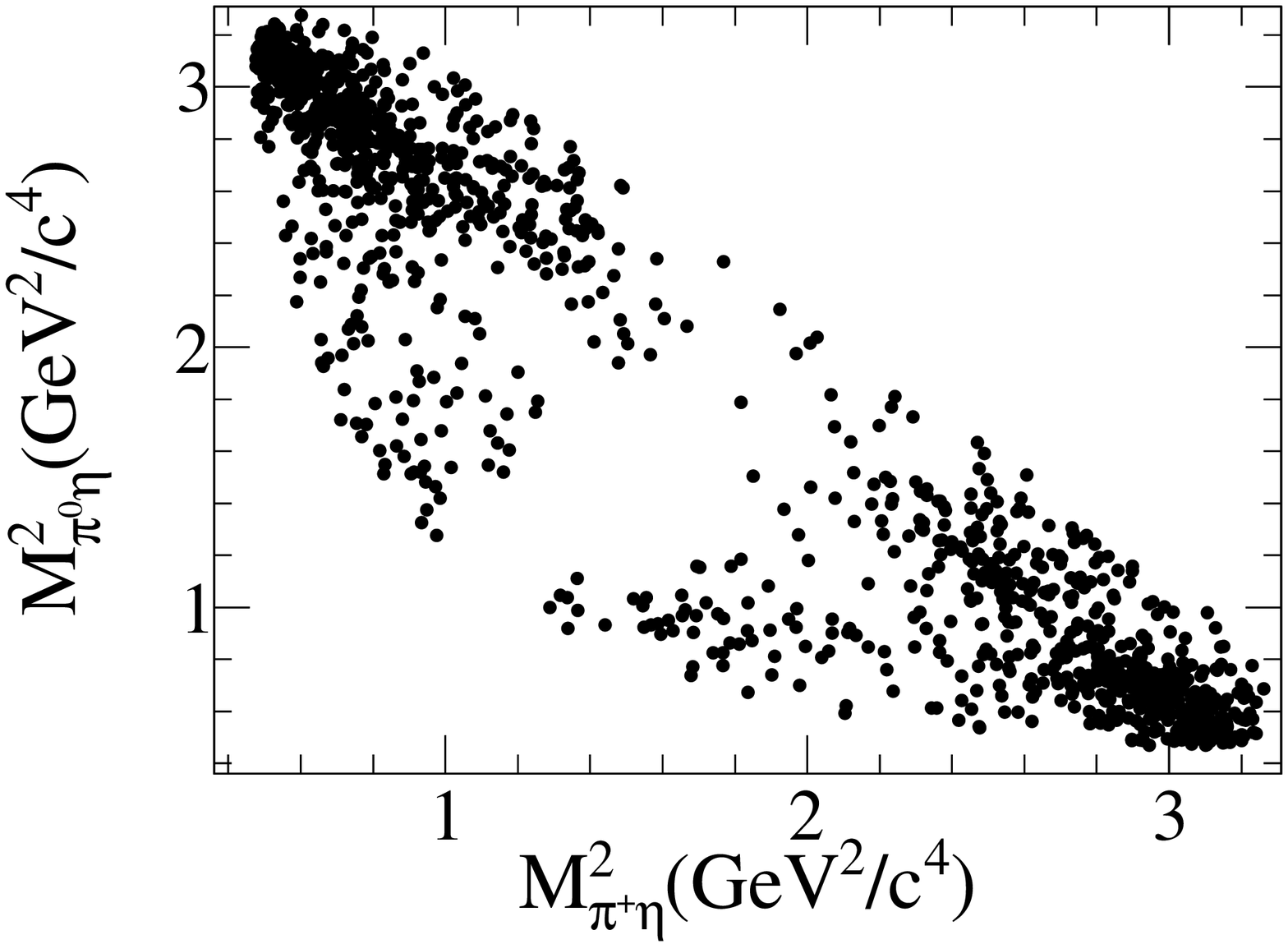}
\put(-35,65){(a)}
\end{minipage}
\begin{minipage}[b]{0.23\textwidth}
\epsfig{width=0.98\textwidth,clip=true,file=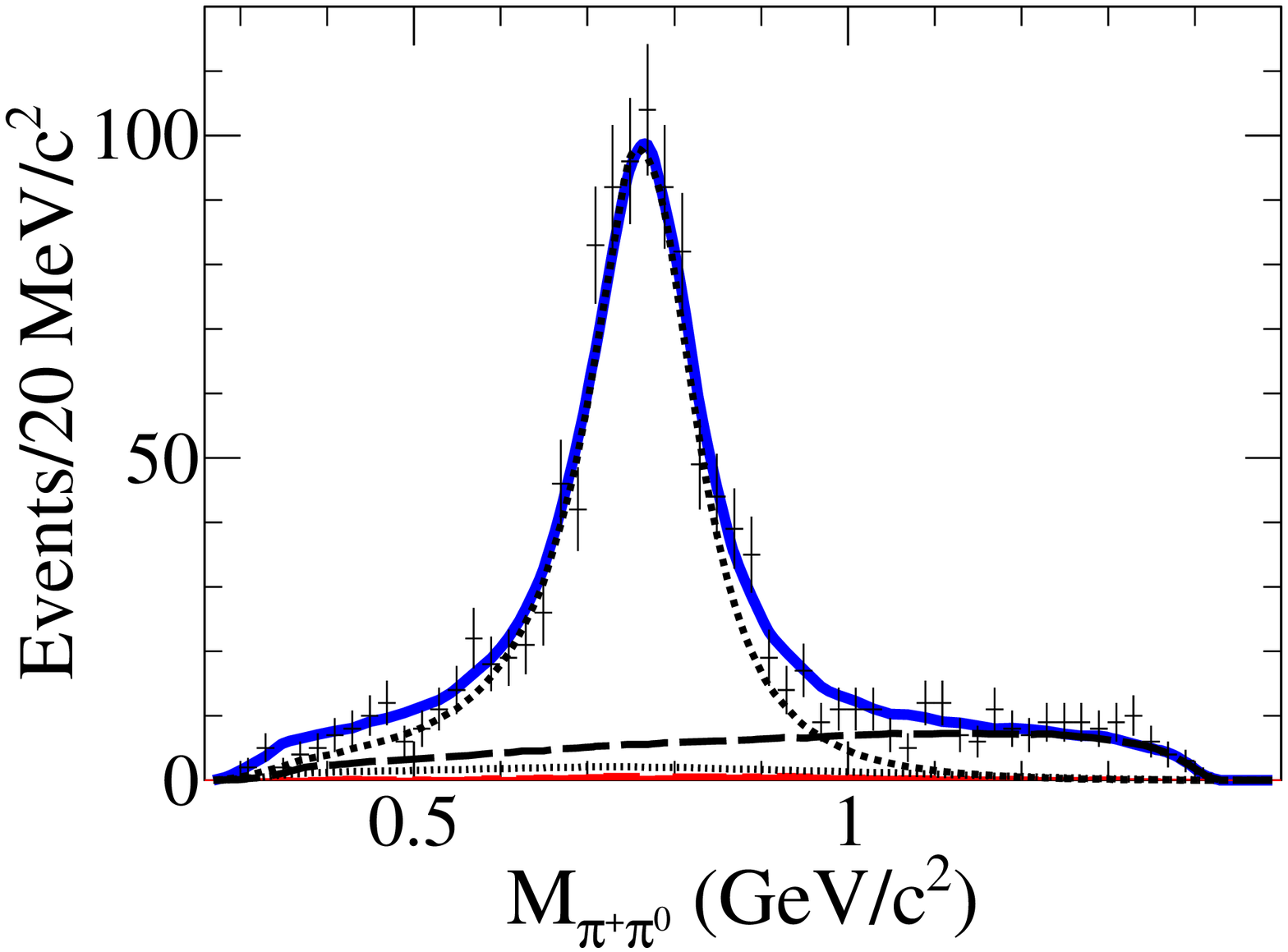}
\put(-35,65){(b)}
\end{minipage}
\begin{minipage}[b]{0.23\textwidth}
\epsfig{width=0.98\textwidth,clip=true,file=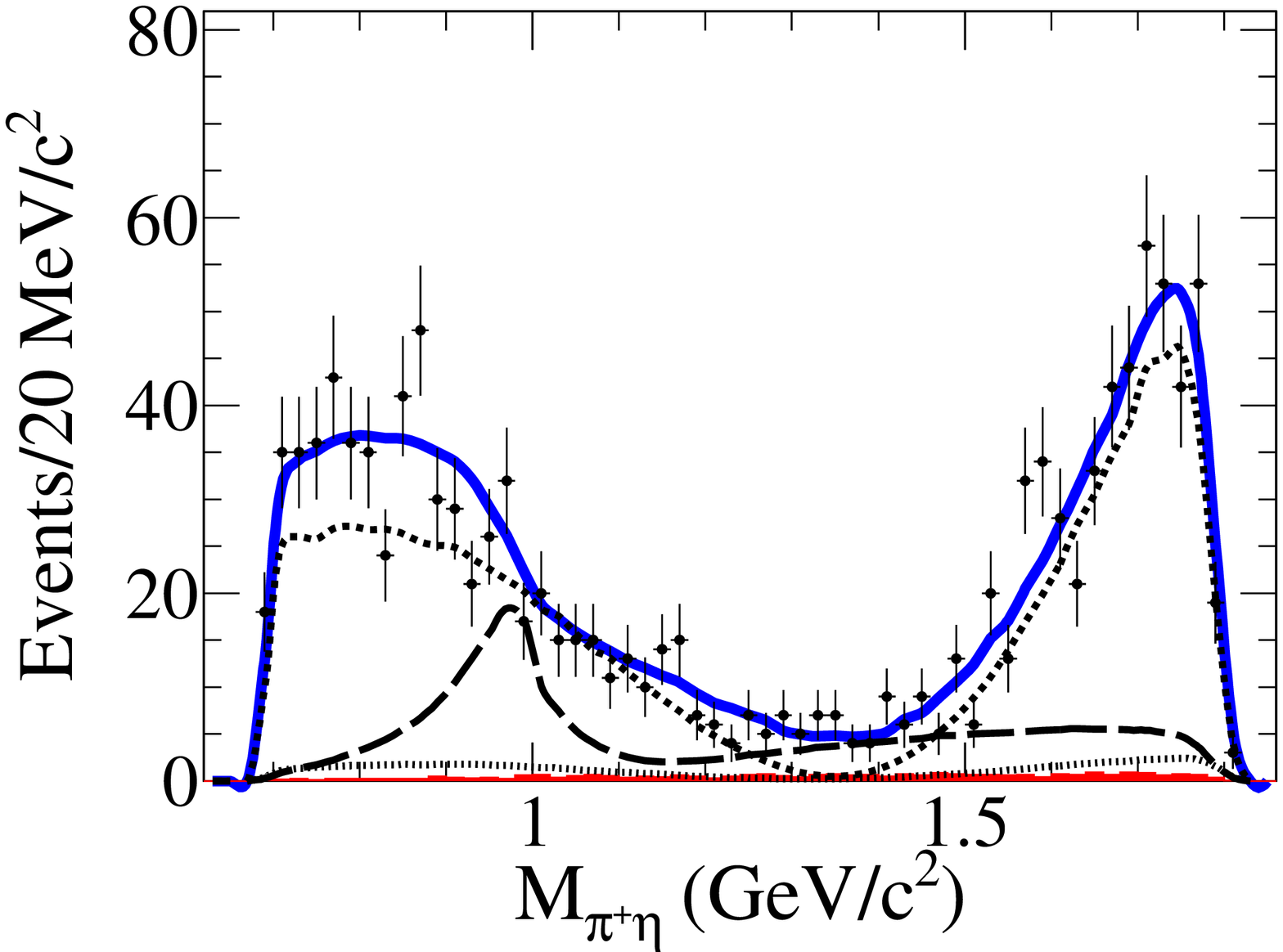}
\put(-35,65){(c)}
\end{minipage}
\begin{minipage}[b]{0.23\textwidth}
\epsfig{width=0.98\textwidth,clip=true,file=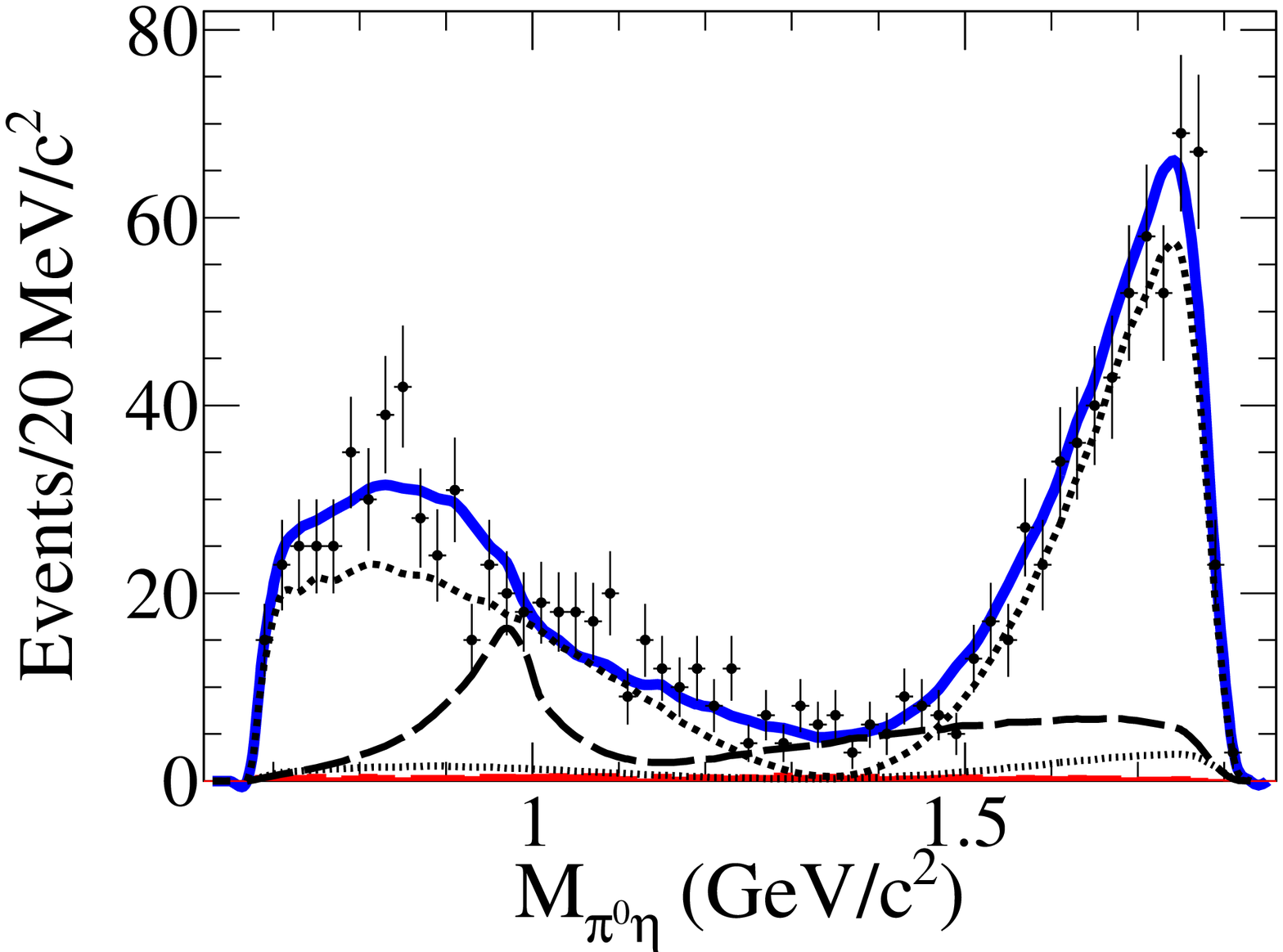}
\put(-35,65){(d)}
\end{minipage}
\begin{minipage}[b]{0.23\textwidth}
\epsfig{width=0.98\textwidth,clip=true,file=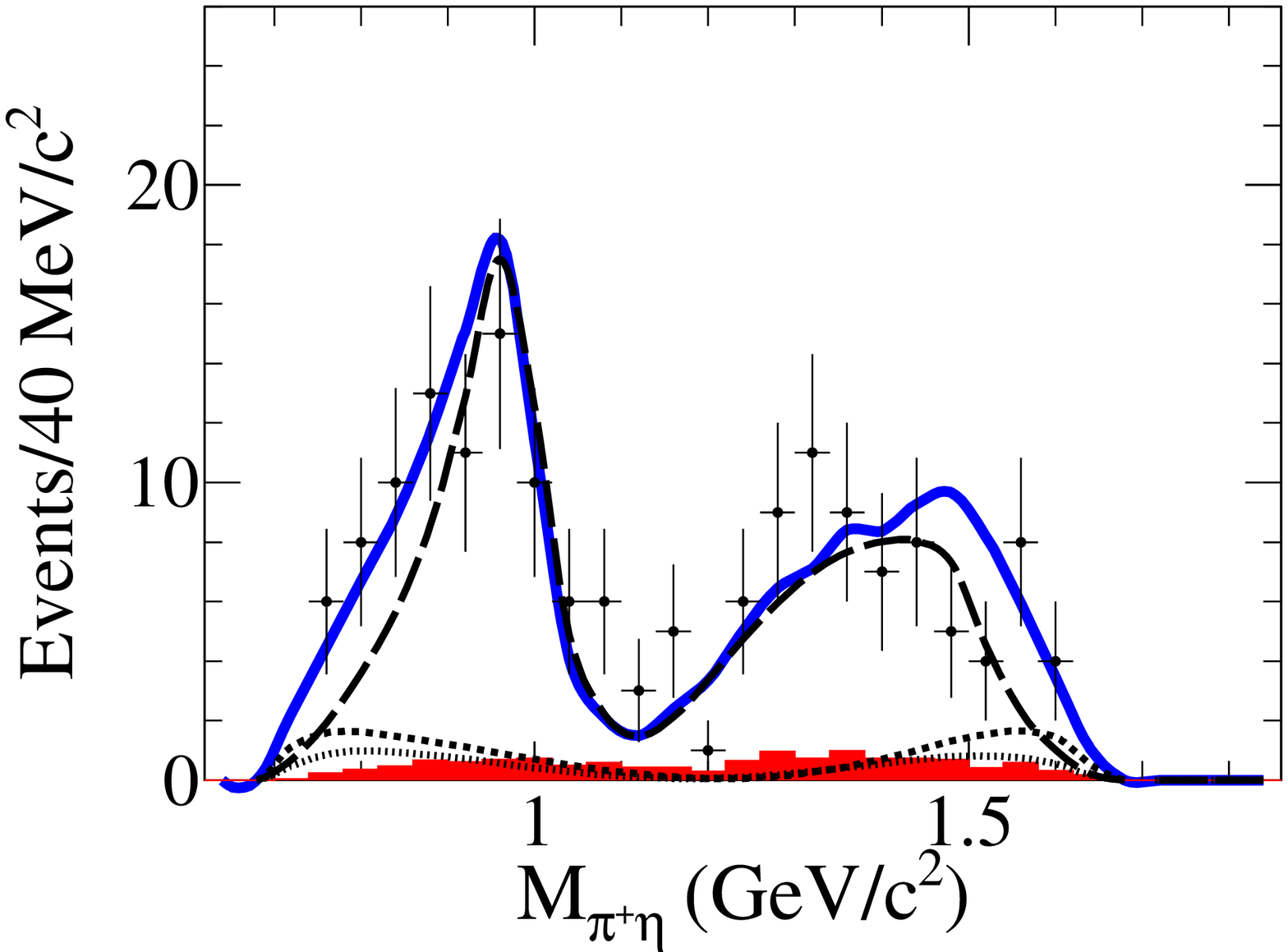}
\put(-35,65){(e)}
\end{minipage}
\begin{minipage}[b]{0.23\textwidth}
\epsfig{width=0.98\textwidth,clip=true,file=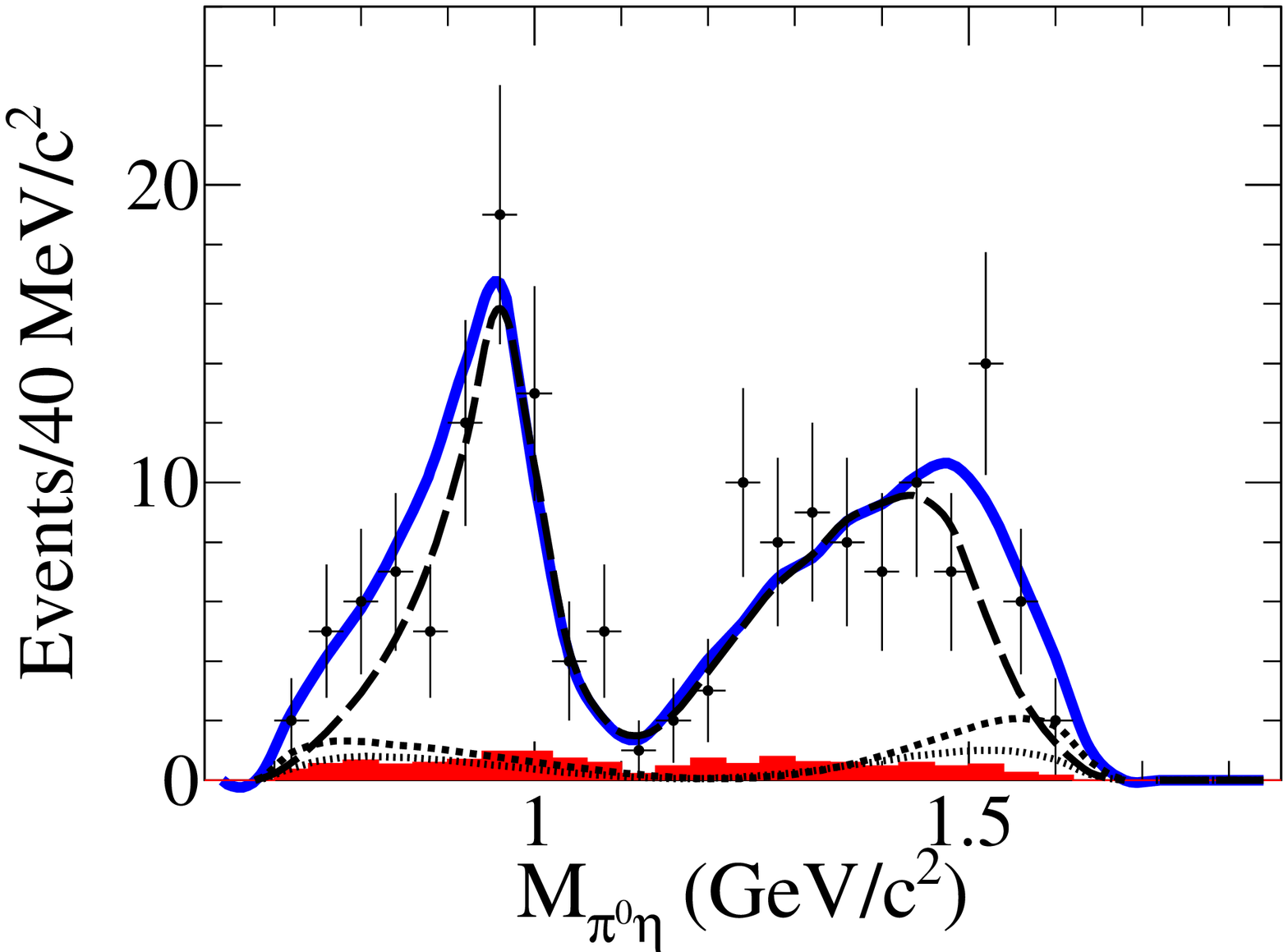}
\put(-35,65){(f)}
\end{minipage}
\caption{(a) Dalitz plot of $M^{2}_{\pi^{+}\eta}$ versus 
$M^{2}_{\pi^{0}\eta}$ for data,  
the projections of the fit on (b) $M_{\pi^{-}\pi^{0}}$, 
(c) $M_{\pi^{+}\eta}$ and (d) $M_{\pi^{0}\eta}$, 
and the projections on (e) $M_{\pi^{+}\eta}$ and (f) $M_{\pi^{0}\eta}$ 
after requiring $M_{\pi^{+}\pi^{0}}>1.0$ GeV$/c^{2}$. 
In (b-f), the dots with error bars and 
the solid line are data and the total fit, respectively; 
the dashed, dotted, and long-dashed  lines are the contributions from $D_{s}^{+} \rightarrow \rho^{+}\eta$, 
$D_{s}^{+} \rightarrow (\pi^{+}\pi^{0})_{V}\eta$, and
$D_{s}^{+} \rightarrow a_{0}(980)\pi$, respectively.  The (red) hatched histograms are the simulated background.}
\label{fig:projection}
\end{center}
\end{figure}

For $D_{s}^{+} \rightarrow \rho^{+}\eta$, $D_{s}^{+} \rightarrow (\pi^{+}\pi^{0})_{V}\eta$, and $D_{s}^{+} \rightarrow a_{0}(980)\pi$, 
the resulting statistical significances are greater than 20$\sigma$, 5.7$\sigma$, and 16.2$\sigma$, respectively. 
Their phases and fit fractions (FFs) are listed in Table~\ref{Tab:result}.
Here the FF for the $n^{{\rm th}}$ intermediate process 
is defined as ${\rm FF}_{n} = \frac{\int{|A_{n}|^{2}d\Phi_{3}}}{\int{|\mathcal{M}|^{2}d\Phi_{3}}}$, where $d\Phi_{3}$ is the standard element of the three-body phase space. 
The Dalitz plot of $M^{2}_{\pi^{+}\eta}$ versus $M^{2}_{\pi^{0}\eta}$ for data is shown in Fig.~\ref{fig:projection}(a).  
The projections of the fit on $M_{\pi^{-}\pi^{0}}$, $M_{\pi^{+}\eta}$ and  $M_{\pi^{0}\eta}$ 
are shown in Figs.~\ref{fig:projection}(b-d). 
The projections on $M_{\pi^{+}\eta}$ and  $M_{\pi^{0}\eta}$ for events with $M_{\pi^{+}\pi^{0}}>1.0$ GeV$/c^{2}$ 
are shown in Figs.~\ref{fig:projection}(e,f), in which $a_{0}(980)$ peaks are observed. 
The fit quality is determined by calculating the $\chi^{2}$ of the fit using an adaptive binning of the $M^{2}_{\pi^{+}\eta}$ versus $M^{2}_{\pi^{0}\eta}$  Dalitz plot that requires each bin contains at least 10 events. 
The goodness of fit is $\chi^{2}$/NDOF=82.8/77. 
\begin{table}[htbp]
\footnotesize
\begin{center}
\caption{Significance, $\phi_n$, and FF$_n$ for the intermediate processes in the nominal fit. The first and second uncertainties are statistical and systematic, respectively.}
\begin{tabular}{ccccc}\hline
Amplitude                                       &$\phi_n$  (rad)        &FF$_n$\\ \hline
$D_{s}^{+} \rightarrow \rho^{+}\eta$            & 0.0 (fixed)           &$0.783\pm0.050\pm0.021$\\
$D_{s}^{+} \rightarrow (\pi^{+}\pi^{0})_{V}\eta$&$0.612\pm0.172\pm0.342$&$0.054\pm0.021\pm0.025$\\
$D_{s}^{+} \rightarrow a_{0}(980)\pi$           &$2.794\pm0.087\pm0.044$&$0.232\pm0.023\pm0.033$\\
\hline
\end{tabular}
\label{Tab:result}
\end{center}
\end{table}

Systematic uncertainties for the amplitude analysis are considered from five sources:
(I) line-shape parameterizations of the resonances, 
(II) fixed parameters in the amplitudes, (III) the background level and distribution in the Dalitz plot, (IV) experimental effects, and (V) the fitter performance. 
We determine these systematic uncertainties separately by taking the difference between 
the values of $\phi_n$, and FF$_n$ found by the altered and nominal fits. 
The uncertainties related to the assumed resonance line-shape are estimated by 
using the following alternatives: a Gounaris-Sakurai function~\cite{GS} for the $\rho^{+}$ 
propagator and a three-channel-coupled Flatt\'e formula, which adds  the $\pi\eta^\prime$ channel~\cite{BAM-168}, for the $a_{0}(980)$ propagator.  
Since varying the propagators results in different normalization factors, 
the effect on all FFs is considered.  The uncertainties related to the fixed parameters in the amplitudes are considered by varying the mass and width of $\rho^{+}$ by $\pm 1\sigma$~\cite{PDG}, the mass and coupling constants of $a_{0}(980)$ by the uncertainties reported in Ref.~\cite{BAM-168}, and the effect of varying the radii of the non-resonant state and $D_{s}$ meson within $\pm 2$~GeV$^{-1}$. In addition, for the $\rho^{+}$ resonance, the effective 
radius reported in Ref.~\cite{PDG} is used as an alternative. 
The uncertainty related to the assumed background level is determined by changing the background fraction within its statistical uncertainty. The uncertainty related to the assumed background shape 
is estimated by using an alternative distribution simulated with  
$D^{+}_{s} \rightarrow \pi^{+}f_{0}(980)$, $f_{0}(980) \rightarrow \pi^{0}\pi^{0}$.  
To estimate the uncertainty from the experimental effect related to the kinematic fits and BDT classifier, 
we set the $\chi^{2}$ requirements for the result of the two kinematic fits to be twice 
the values used in the nominal selection, alter the $\cos \theta_{\eta}$ requirement to be greater than 0.996, and adjust the BDT requirement such that the purity is approximately equal to the sample used in the nominal fit. The fitter performance is investigated with the same method as reported in Ref.~\cite{K3Pi}. The biases are small and taken as the systematic uncertainties.
The contributions of individual systematic uncertainties are summarized in Table~\ref{Tab:sys_unc}, 
and are added in quadrature to obtain the total systematic uncertainty.

\begin{table}[htbp]
\begin{center}
\caption{Systematic uncertainties on the $\phi$ and FFs for different amplitudes 
in units of the corresponding statistical uncertainties.}
\begin{tabular}{ccccccccccccc}\hline
\multirow{2}{*}{Amplitude}                      &$~$   &\multicolumn{5}{c}{Source}& $~$\\
                                                &$~$   &I   &II  &III &IV  & V  &Total\\ \hline
$D_{s}^{+} \rightarrow \rho^{+} \eta$           &FF    &0.06&0.34&0.13&0.12&0.15&0.41 \\
\hline
\multirow{2}{*}{$D_{s}^{+} \rightarrow (\pi^{+}\pi^{0})_{V}\eta$}
                                                &$\phi$& $-$&1.97&0.18&0.03&0.17&1.99 \\
                                                &FF    &0.61&1.03&0.12&0.06&0.08&1.21 \\
\hline
\multirow{2}{*}{$D_{s}^{+} \rightarrow a_{0}(980) \pi$}
                                                &$\phi$& $-$&0.41&0.07&0.28&0.09&0.51 \\
                                                &FF    &0.58&1.31&0.02&0.06&0.11&1.45 \\
\hline
\end{tabular}
\label{Tab:sys_unc}
\end{center}
\end{table}

\begin{figure}[tp]
\begin{center}
\begin{minipage}[b]{0.48\textwidth}
\epsfig{width=0.98\textwidth,clip=true,file=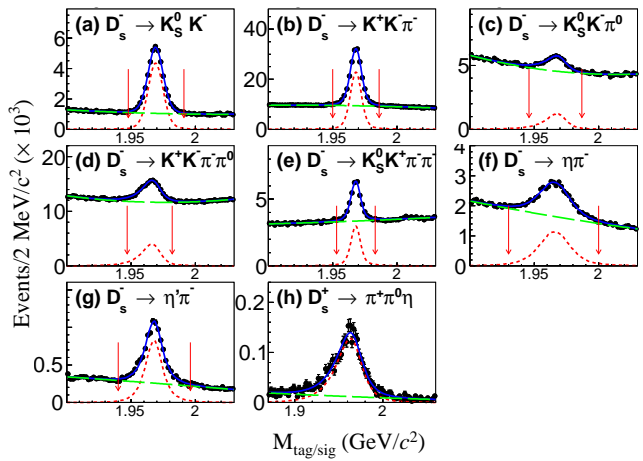}
\end{minipage}
\caption{Fits to (a-g) the $M_{{\rm tag}}$ distributions of seven tag modes (indicated  in each sub-figure) and (h) the $M_{{\rm sig}}$ distribution of signal candidates.
The dots with error bars are data. 
The (blue) solid lines are the total fit.
The (red) dashed and the (green) long-dashed lines are signal and background, respectively. 
In (a-g), the $D_{s}^{-}$ signal regions are between the arrows.}
\label{fig:DT}
\end{center}
\end{figure}

Further, we measure the total BF of $D_{s}^{+} \rightarrow \pi^{+}\pi^{0}\eta$ without reconstructing $\gamma_{\rm direct}$ to improve the statistical precision. 
The ST yields ($Y_{{\rm tag}}$) and DT yield ($Y_{{\rm sig}}$) of data 
are determined by the fits to the resulting $M_{{\rm tag}}$ and $M_{{\rm sig}}$ distributions, 
as shown in Figs.~\ref{fig:DT}(a-g) and Fig.~\ref{fig:DT}(h), respectively. 
In each fit, the signal shape is modeled with the MC-simulated shape convoluted with a Gaussian
function, which accounts for any difference in resolution between the data and MC, and the background is described with a second-order Chebychev polynomial. 
These fits give a total ST yield of $Y_{{\rm tag}}=255895\pm1358$ and a signal yield of $Y_{{\rm sig}}=2626\pm77$.
Based on the signal MC sample, generated according to the amplitude analysis results reported in this Letter, the DT efficiencies ($\epsilon_{{\rm tag,sig}}$) are determined. 
With $Y_{{\rm tag}}$, $Y_{{\rm sig}}$, $\epsilon_{{\rm tag,sig}}$ and the ST efficiencies ($\epsilon_{{\rm tag}}$),  
the relationship 
$\mathcal{B}(D_{s}^{+} \rightarrow \pi^{+}\pi^{0}\eta) = \frac{Y_{{\rm sig}}}{\sum_{i}{Y_{{\rm tag}}^{i}
\epsilon_{{\rm tag,sig}}^{i}/\epsilon_{{\rm tag}}^{i}}}$, where the index $i$ denotes the $i^{{\rm th}}$ tag mode, 
is used to obtain $\mathcal{B}(D_{s}^{+} \rightarrow \pi^{+}\pi^{0}\eta) = (9.50\pm0.28_{{\rm stat.}})\%$. 

For the total BF measurement, the systematic uncertainty related to the signal shape is studied
by performing an alternative fit without convolving the Gaussian resolution function. The BF shift of 0.5\% is taken as the uncertainty.
The systematic uncertainty arising from the assumed background shape and 
the fit range is studied by replacing our nominal ones with a  
first-order Chebychev polynomial and a fit range of 
$[1.88,\,2.04]$ GeV$/c^{2}$, respectively. The largest BF shift of 0.6\% is taken as the related uncertainty.
The possible bias due to the measurement method 
is estimated to be 0.2\% by comparing the measured BF in the GMC sample, using the same method as in data analysis, to the value assumed in the generation. 
The uncertainties from particle identification and tracking efficiencies are assigned to be  
0.5\% and 1.0\%~\cite{omegapi}, respectively.
The relative uncertainty in the $\pi^{0}$ reconstruction efficiency is 2.0\%~\cite{omegapi}, and the uncertainty in 
$\eta$ reconstruction is assumed to be comparable to that on $\pi^{0}$ reconstruction and correlated with it.
The uncertainty from the Dalitz model of 0.6\% is estimated as the change of efficiency when 
the model parameters are varied by their systematic uncertainties 
(this term is not considered when calculating the BFs of the intermediate processes).
The uncertainties due to MC statistics (0.2\%) and the value of 
$\mathcal{B}(\pi^{0}/\eta \rightarrow \gamma\gamma)$ used \cite{PDG} (0.5\%) are also considered. 
Adding these uncertainties in quadrature gives  
a total systematic uncertainty of 4.3\%.

We obtain $\mathcal{B}(D_{s}^{+} \rightarrow \pi^{+}\pi^{0}\eta)$ to be 
$(9.50\pm0.28_{{\rm stat.}}\pm0.41_{{\rm sys.}})\%$. Using the FFs listed in Table~\ref{Tab:result},  
the BFs for the intermediate processes
$D_{s}^{+} \rightarrow \rho^{+}\eta$ and $D_{s}^{+} \rightarrow (\pi^{+}\pi^{0})_{V}\eta$
are calculated to be $(7.44\pm0.52_{{\rm stat.}}\pm0.38_{{\rm sys.}})$\% and 
$(0.51\pm0.20_{{\rm stat.}}\pm0.25_{{\rm sys.}})$\%, respectively.
With the definition of fit fraction, fraction of $D_{s}^{+} \rightarrow a_{0}(980)^{+(0)}\pi^{0(+)}, a_{0}(980)^{+(0)} \to \pi^{+(0)} \eta$
with respect to the total fraction of $D_{s}^{+} \rightarrow a_{0}(980)\pi, a_{0}(980) \to \pi \eta$ is evaluated to be 0.66. 
Multiplying by the FF of $D_{s}^{+} \rightarrow a_{0}(980)\pi$ determined from the nominal fit and $\mathcal{B}(D_{s}^{+} \rightarrow \pi^{+}\pi^{0}\eta)$, 
the BF of $D_{s}^{+} \rightarrow a_{0}(980)^{+(0)}\pi^{0(+)}, a_{0}(980)^{+(0)} \to \pi^{+(0)} \eta$ is determined to be 
$(1.46\pm0.15_{{\rm stat.}}\pm0.23_{{\rm sys.}})$\%.
 
In summary, we present the first amplitude analysis of the decay $D_{s}^{+}\rightarrow \pi^{+}\pi^{0}\eta$. 
The absolute BF of $D_{s}^{+}\rightarrow \pi^{+}\pi^{0}\eta$ is measured 
with a precision improved by a factor of 2.5 compared with the world average value~\cite{PDG}.  
We observe the pure WA decays 
$D_{s}^{+} \rightarrow a_{0}(980)\pi$ for the first time 
with a statistical significance of 16.2$\sigma$. 
The measured $\mathcal{B}(D_{s}^{+} \rightarrow a_{0}(980)^{+(0)}\pi^{0(+)})$ 
is larger than other measured BFs of pure WA decays 
$D_{s}^{+} \rightarrow \omega \pi^{+}$ and $D_{s}^{+} \rightarrow \rho^{0} \pi^{+}$ by at least one order of magnitude.
Furthermore, when the measured $a_{0}(980)^{0}$-$f_{0}(980)$ mixing rate~\cite{Ablikim:2018pik} is considered,
the expected effect of $a_{0}(980)^{0}$-$f_{0}(980)$ mixing is lower than the WA contribution in $D_{s}^{+} \rightarrow a_{0}(980)^{0}\pi^{+}$ decay 
by two orders of magnitude, which is negligible.
 
With the measured $\mathcal{B}(D_{s}^{+} \rightarrow a_{0}(980)^{+(0)}\pi^{0(+)})$, 
the WA contribution with respect to the tree-external-emission contribution in $SP$ mode is estimated to be $0.84\pm0.23$~\cite{AoverT},
which is significantly greater than that (0.1$\sim$0.2) in $VP$ and $PP$ modes~\cite{Li:2013xsa,HaiYangCheng2}. 
This measurement sheds light on the FSI effect and non-perturbative strong interaction~\cite{HaiYangCheng1,HaiYangCheng2}, and
provides a theoretical challenge to understand such a large WA contribution.
In addition, the result of this analysis is an essential input to determine the effect from $a_{0}(980)^{0}$ on the $K^{+}K^{-}$ 
$S$-wave contribution to the model-dependent amplitude analysis of $D_{s}^{+} \rightarrow K^{+}K^{-}\pi^{+}$~\cite{Mitchell:2009aa,delAmoSanchez:2010yp}.

The authors greatly thanks Prof. Fu-Sheng Yu and Prof. Haiyang Cheng for the useful discussions.
The BESIII collaboration thanks the staff of BEPCII and the IHEP
computing center for their strong support.
This work is supported in part by National Key Basic Research Program of China
under Contract No. 2015CB856700; National Natural Science Foundation of China (NSFC)
under Contracts Nos. 11475185, 11625523, 11635010, 11735014;
the Chinese Academy of Sciences (CAS) Large-Scale Scientific Facility Program;
the CAS Center for Excellence in Particle Physics (CCEPP);
Joint Large-Scale Scientific Facility Funds of the NSFC and CAS
under Contracts Nos. U1332201, U1532257, U1532258;
CAS Key Research Program of Frontier Sciences under Contracts
Nos. QYZDJ-SSW-SLH003, QYZDJ-SSW-SLH040;
100 Talents Program of CAS; National 1000 Talents Program of China;
INPAC and Shanghai Key Laboratory for Particle Physics and Cosmology;
German Research Foundation DFG under Contracts Nos. Collaborative Research Center CRC 1044, FOR 2359;
Istituto Nazionale di Fisica Nucleare, Italy;
Koninklijke Nederlandse Akademie van Wetenschappen (KNAW) under Contract No. 530-4CDP03;
Ministry of Development of Turkey under Contract No. DPT2006K-120470;
National Science and Technology fund;
The Swedish Research Council;
U. S. Department of Energy under Contracts Nos. DE-FG02-05ER41374, DE-SC-0010118, DE-SC-0010504, DE-SC-0012069; University of Groningen (RuG) and the Helmholtzzentrum fuer Schwerionenforschung GmbH (GSI), Darmstadt;
WCU Program of National Research Foundation of Korea under Contract No. R32-2008-000-10155-0.

\end{document}